\documentclass[graybox]{svmult}

\usepackage{type1cm}        
\usepackage{makeidx}         
\usepackage{graphicx}        
\usepackage{multicol}        
\usepackage[bottom]{footmisc}

\usepackage{newtxtext}       %
\usepackage[varvw]{newtxmath}       

\makeindex             

\begin{document}

\title*{Radio Galaxy Zoo: EMU - paving the way for EMU cataloging using AI and citizen science}
\titlerunning{Radio Galaxy Zoo: EMU - AI and Citizen Science for Cataloging}
\author{Hongming Tang, Eleni Vardoulaki and RGZ EMU collaboration}
\institute{Hongming Tang \at Department of Physics, Xi'an Jiaotong-Liverpool University, Suzhou 215123, China, \email{Hongming.Tang@xjtlu.edu.cn}
\and Eleni Vardoulaki \at IAASARS, National Observatory Athens, Lofos Nymfon, 11851 Athens, Greece \email{elenivard@gmail.com}}
%
%
\maketitle

\abstract{The Evolutionary Map of the Universe (EMU) survey with ASKAP is transforming our understanding of radio galaxies, AGN duty cycles, and cosmic structure. EMUCAT efficiently identifies compact radio sources, yet struggles with extended objects, requiring alternative approaches. The Radio Galaxy Zoo: EMU (RGZ EMU) project proposes a general framework that combines citizen science and machine learning to identify $\sim$ 4 million extended sources in EMU. This framework is expected to enhance the EMUCAT cataloging on extended sources and can be further empowered with the introduction of cross-matched external data from surveys such as POSSUM and WALLABY.}

\section{Overview}
\label{sec:1}

The Evolutionary Map of the Universe (EMU) project, using the Australian Square Kilometre Array Pathfinder (ASKAP) radio telescope, surveys the southern hemisphere at $\sim$1 GHz to advance our understanding of radio galaxy astrophysics, AGN duty cycles, cosmic star formation history, galaxy cluster mergers, and unexpected discoveries (\cite{norris_emu}, Hopkins et al., submitted). The cataloging workflow is managed by EMUCAT, a system designed to integrate ASKAP observations into cohesive data products, incorporating multi-waveband data and performing product associations. Although EMUCAT handles most unresolved objects well ($\sim$80\% of EMU sources), it produces less reliable results for extended sources such as radio galaxies, star-forming galaxies, and unusual sources such as radio relics, galaxy cluster bridges, and Odd Radio Circles (ORCs; \cite{norris_orc}).

To address these challenges, additional methods are employed that include expert visual inspection, machine learning, and citizen science. Visual inspection of EMU pilot survey data (Yew et al., in prep.) has played a crucial role in training RG-CAT \cite{gupta_rgcat}, a comprehensive pipeline designed to identify both compact and extended radio sources. It has made significant progress in identifying extended sources and cross-matching their multi-waveband counterparts, though there remains space for improvement, particularly with complex tasks. To maximize our ability to identify these morphologically complex or unusual objects, fuel the active learning iterations of machine learning based pipelines like RG-CAT and more downstream tasks, we design a cataloging framework that combines machine learning and citizen science.

\section{AI driven sample selection and project designation}

\begin{figure}
\includegraphics[width=0.9\textwidth]{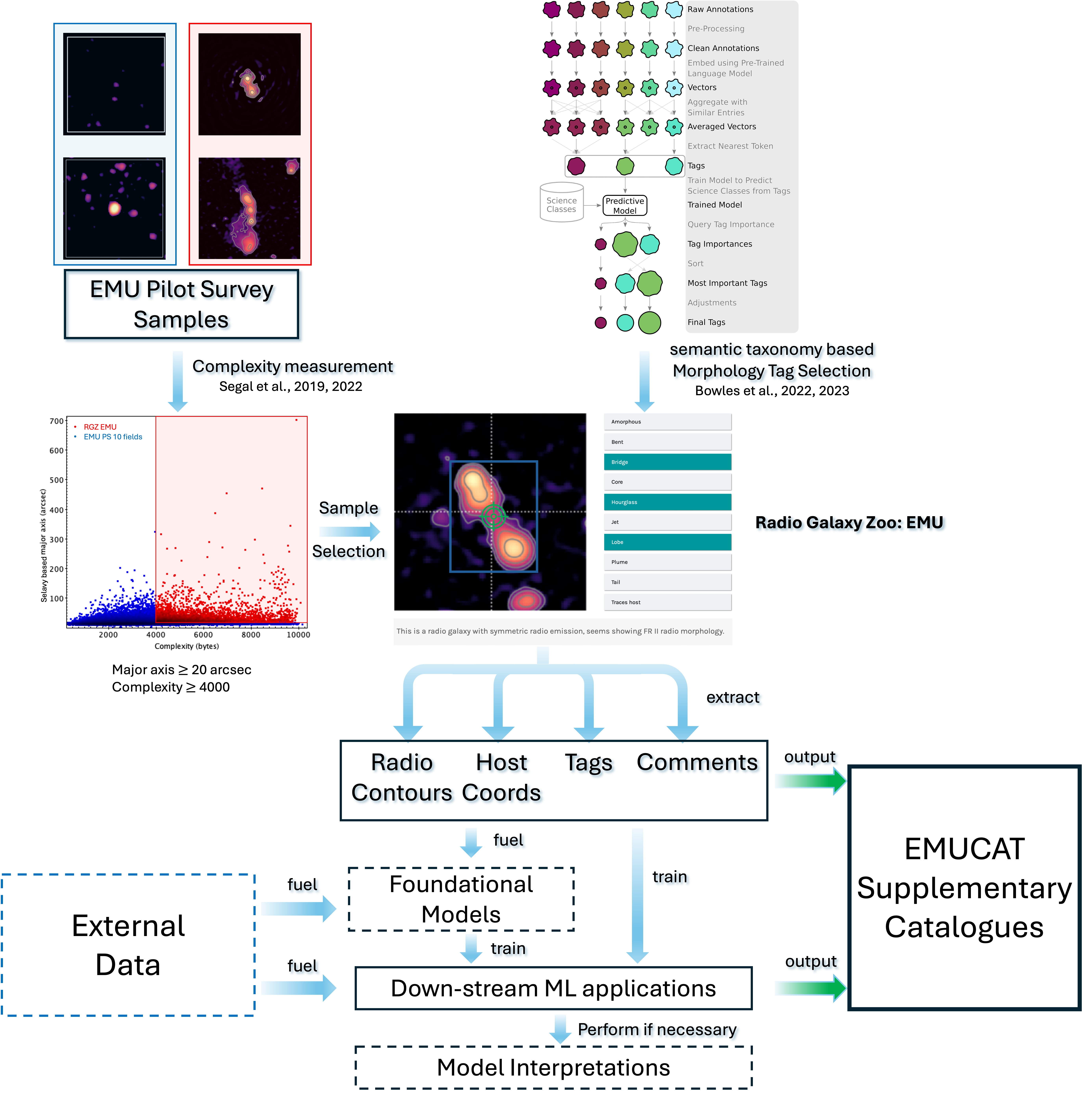}
%
%
\caption{A schematic diagram showing the proposed RGZ EMU cataloging framework using citizen science and machine learning. The blue box (top left panel) shows sample images that are excluded from the final selection, as having either a low complexity or a source major axis smaller than 20 arcsec. Sample images in the red box, on the contrary, refers to sample images fulfilled our selection criteria. Gradient blue arrows refers to operation done within the framework, and gradient green arrows indicates the framework components that will contribute to the EMUCAT supplementary catalogs.}
\label{fig:workflow}       
\end{figure}

Through the 5.5-year operation of Radio Galaxy Zoo \cite{wong_rgzdr1}, we recognized that even with the support of around 12,000 RGZ volunteers, it would takes us $\sim$ 157 years to handle the identification task of $\sim$ 4 million extended radio sources expected to be discovered by EMU. To develop a quasi-automated framework for EMU extended source cataloging, the combination of both citizen science and machine learning hence becomes necessary. In this section, we introduce our AI+citizen science framework, where Radio Galaxy Zoo: EMU project (RGZ EMU hereafter) will serve as its ``nerve center'' (Figure~\ref{fig:workflow}).

\subsection{Sample Selection}
\label{sec:sample_selection}

To maximize the scientific value and efficiency of such a framework, a representative, diverse, and scientifically meaningful sample should be selected to feed our citizen science project as their identification results will not only become part of the EMU value-added catalog but fuel the training and iteration of a series of downstream machine learning algorithms. 

Starting from more than 200,000 entries from the EMU PS 1 Selavy component catalog, we created their 6' by 6' image cutouts and measured their cutout ``complexity'', a metric developed to estimate the quantity of information embedded in a given image (see \cite{segal_complexity} for detailed explanation). With our scientific goal in mind, we set a lower complexity limit of 4,000 and reduced the sample number to 37,578. In order to ensure that the central targeted source in an image shows extended morphology, we further required that an entry can only be considered if its Selavy-derived component major axis has an angular size larger than 20 arcsec. With both criteria applied, we selected 6230 sample images for the Phase 1 operation of the RGZ EMU.

\subsection{Radio Morphology Tags}

Our current radio galaxy classification system has grown substantially, which has led to challenges in maintaining consistency and organization \cite{rudnick_tagnotbox}. It was then suggested to allow people to assign multiple tags to a source. Motivated by this idea, we investigated the possibility of asking volunteers to tag the radio morphology of a source. 

For citizen science to be feasible, we need to bridge the knowledge gap between radio astronomers and citizen scientists. This is achieved by implementing a novel semantic taxonomy-based natural language processing approach, developing an adaptable English taxonomy for radio astronomy, and finally deriving 22 semantic tags capable of describing source radio morphology \cite{bowles_taxonomy}. Moreover, to enhance scientific results and streamline decision-making, since an excessive number of choices can increase complexity \cite{Iyengar_gooddecision}, we assessed the difficulty of algorithmically computing tags at different stages of data processing and reduced the number of tags to 10 \cite{bowles_taxonomy}.

\section{Citizen Science driven machine learning applications}

\begin{figure}
\includegraphics[width=0.9\textwidth]{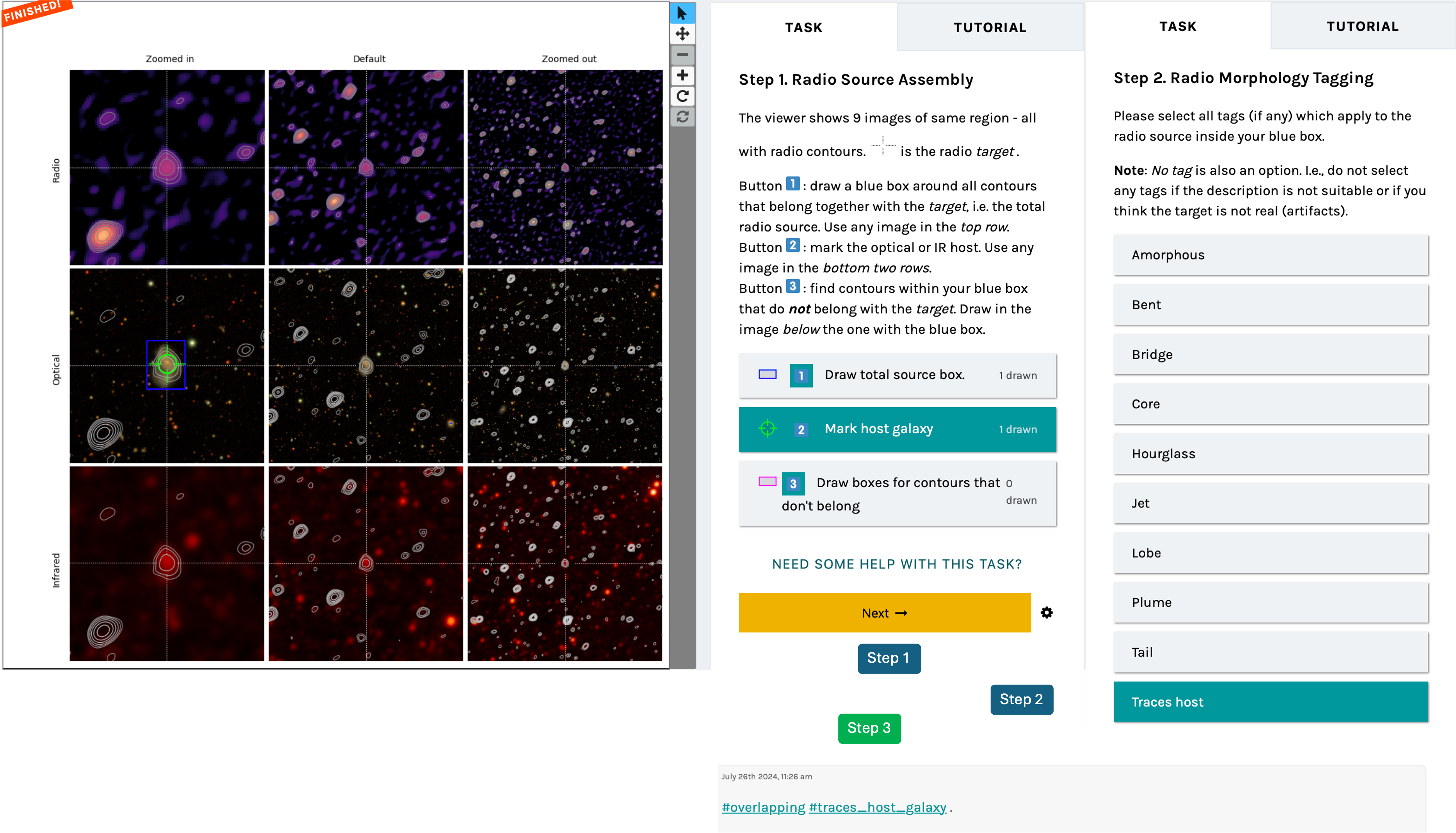}
%
%
\caption{RGZ EMU user interface and its 3-step workflow setup. The first 2 steps (in blue color) of the workflow is compulsary, while the third (in green color) is optional.}
\label{fig:user_interface}       
\end{figure}

With the sample data and tag options settled down, we are able to design and setup the RGZ EMU. Figure~\ref{fig:user_interface} shows the RGZ EMU user interface (UI) and its embedded functionalities / tasks that we built for citizen scientists.  

\begin{description}[Type 1]
\item[1. Radio Source Assembly:]{associate source radio emission regions and their optical/IR counterparts.}
\item[2. Radio Morphology Tagging:]{identify source radio morphology using tags}
\item[3. Talkboard:]{describe radio source properties, alarm sources with unusual or unexpected morphology}
\end{description}

We can see from Figure~\ref{fig:user_interface} that volunteers are asked to work on the tasks above by looking at sample radio data from the ASKAP EMU pilot survey, infrared data from $WISE$ 3.4$\rm \mu m$ observations and optical data from Dark Energy Survey (DES) at different angular scale (3' $\times$ 3', 6' $\times$ 6', and 12' $\times$ 12'). RGZ EMU task output data (i.e., radio contours, optical/IR host coordinates, morphology tags, subject-wise comments, or corpus) would eventually be included in the EMUCAT supplementary catalog after some automated data processing.

Once equipped with the resulting data comprising images, regions, coordinates, tags, and corpus, we can evaluate the data quality by comparing the classifications of volunteers and experts on a small set of image samples known as golden samples. We will then utilize data exceeding the given consensus level to develop and refine a series of machine learning algorithms at both the foundational and downstream levels. Building on our previous RGZ experience and the need for trustworthy AI in the SKA era, we argue that a framework combining foundational models, downstream models with specific objectives, and methods for explaining model prediction rationales can facilitate the creation of a scientifically valuable and trustworthy AI-driven EMUCAT supplementary catalog. In line with the vision of EMU, the scientific value of such a framework can only be enhanced when our data are matched with other surveys, such as POSSUM, PEGASUS, WALLABY, and {\it Euclid}. 

\section{Discussion}

Since its launch on 8 July 2024, RGZ EMU has attracted 1,435 citizen scientists around the world and completed over 53,000 classifications. The project has been translated into 4 languages (English, Greek, Urdu, and Chinese) and can be accessed via Zooniverse, SKA Observatoty (SKAO) and China National Astronomical Data Center (NADC). We expect to release early cataloging results and machine learning applications in upcoming publications by our collaboration members.

\begin{acknowledgement}
HMT gratefully acknowledges the grant support from the Shuimu Tsinghua Scholar Program of Tsinghua University, the 2025 IAU Office of Astronomy for Development (IAU OAD); HMT also acknowledges support from the Department of Physics of Xi'an Jiaotong-Liverpool University, the long-lasting support from the JBCA machine learning group, DoA Tsinghua TAGLAB and machine learning group. 
\end{acknowledgement}

%
%

\end{document}